\documentclass[aps,pra,showpacs,twocolumn]{revtex4}
\usepackage{amsmath}
\usepackage{amssymb}
\usepackage{epsfig}
\usepackage{color}

\graphicspath{{.}{./EPS/}}

\newcommand* {\bra}[1]{\ensuremath{\langle {#1} |}}
\newcommand* {\ket}[1]{\ensuremath{| {#1} \rangle}}
\newcommand* {\ee}{\ensuremath{\mathrm{e}}}

\begin{document}
 
\title{Photon-assisted entanglement creation by minimum-error generalized quantum  measurements in the strong coupling regime}
\author{J. Z. Bern\'ad}
\affiliation{Institut f\"{u}r Angewandte Physik, Technische Universit\"{a}t Darmstadt, D-64289, Germany}
\author{G. Alber}
\affiliation{Institut f\"{u}r Angewandte Physik, Technische Universit\"{a}t Darmstadt, D-64289, Germany}

\date{\today}

\begin{abstract}
We explore possibilities of entangling two distant material qubits with the help of an optical radiation
field in the regime of strong quantum electrodynamical coupling with almost resonant interaction.
For this purpose the optimum generalized field measurements are determined which are capable of preparing
a two-qubit Bell state by postselection with minimum error.
It is demonstrated that
in the strong-coupling regime
some of the recently found limitations of the non-resonant weak-coupling regime can be circumvented successfully
due to characteristic quantum electrodynamical quantum interference effects.
In particular, in the absence of photon loss it is possible to postselect two-qubit
Bell states with fidelities close to unity by a proper choice of the
relevant interaction time. 
Even in the presence of photon loss this strong-coupling regime offers interesting perspectives for creating
spatially well-separated Bell
pairs with high fidelities, high success probabilities,
and high repetition rates which are
relevant for future realizations of quantum repeaters.
\end{abstract}


\maketitle

\section{Introduction}
Entanglement is a primary resource for quantum technology \cite{Leuchs}. For 
applications in quantum communication, such as quantum key distribution, for example, the generation of well-controlled
entanglement between spatially separated quantum systems is of crucial importance. 
For this purpose quantum repeaters \cite{repeaterreview}
are needed which counteract the destructive influence of uncontrolled environmental interactions.

Since the early work of Briegel et al.\cite{Briegel98,Duer99}
there have been numerous theoretical proposals suggesting different physical platforms 
for realizing quantum repeaters \cite{repeaterreview}.
They are based on the main idea of creating entanglement between quantum systems over large
distances with the help of a chain of many uncorrelated
pairs of quantum systems each of which is entangled over a significantly shorter distance only. By performing appropriate
Bell measurements on each of the two qubits of adjacent entangled pairs 
it is possible to swap the already existing short-distance entanglement to
the far-separated outermost quantum systems of such a chain.

From the experimental point of view the realization of a quantum repeater still constitutes a major
technological challenge. Essential for any such realization are two prerequisites, namely
efficient
physical mechanisms for generating highly entangled pairs of quantum systems
with high success probabilities and high repetition rates
and optimal ways for implementing complete Bell measurements accurately.
It has been demonstrated theoretically \cite{repeaterreview} that the exchange of photons
provides a powerful means for entangling
material quantum systems at least over distances of moderate lengths, say a few kilometers, thus
suggesting practicable solutions for the first prerequisite. 

An interesting example in this respect is
the recent theoretical
proposal of van Loock et al. \cite{vanLoock1,vanLoock2,vanLoock3} of a hybrid
quantum repeater based on continuous variables. 
It suggests the exchange of a
single-mode coherent state
of an optical radiation field  between two cavities
for the purpose of entangling spatially separated material qubits.
It takes advantage of the fact that experimentally
these field states can be controlled well and that they can also be produced
with high repetition rates.
The main idea of this proposal is to
entangle this optical radiation field 
with two initially
uncorrelated material qubit systems quantum electrodynamically
and to create entanglement between these two
qubits by an appropriate measurement of the quantum state of the radiation field which postselects a Bell state of
the two material qubits. 
In their original proposal van Loock et al. \cite{vanLoock1,vanLoock2}
discuss cases in which the qubit systems couple to the radiation field 
in a non-resonant way inside two spatially separated cavities connected by an optical fiber
so that their quantum electrodynamical interaction is weak and can be described perturbatively. 
Although offering
a transparent theoretical description this perturbative regime of the electromagnetic coupling
also causes major theoretical limitations as far as 
the achievable degree of entanglement between the material qubits
is concerned. They can be traced back to the fact that the relevant field states
which have to be distinguished in order to postselect a Bell state of the
material qubit pair are not orthogonal.
Thus, 
these field states
cannot be distinguished perfectly by any quantum measurement so that the entanglement of the
postselected material two-qubit state is never perfect.

The basic ideas of
this theoretical proposal offer interesting perspectives for the physical realization of entanglement sources.
In view of these developments 
the natural question arises whether it is possible
to circumvent the theoretical limitations of the weak coupling regime
and to provide a physical mechanism which is capable of
producing entangled two-qubit
pairs not only with a high repetition rate and high success probability but also with arbitrarily
high degree of entanglement. For any future realization of a quantum repeater such a 
mechanism for creating at least short- to intermediate-distance entanglement between two qubits
is useful
as it is expected to increase final rates of producing
long-distance entanglement by subsequent entanglement 
swapping and quantum state purification
significantly. It is a main aim of this paper to address this question.

In the following it is demonstrated that the strong coupling limit of the quantum electrodynamical interaction offers
interesting perspectives for photon-assisted entanglement creation between material quantum systems.
Coupling two distant
material few-level systems almost resonantly to 
the quantized radiation field
the
performance
of 
entanglement creation processes, such as the one originally proposed by van Loock et al. \cite{vanLoock1},
can be improved significantly. 
This way it is possible to circumvent
previously discussed theoretical limitations 
which result from the restriction 
of the quantum electrodynamical interaction to the weak coupling limit.
In contrast, in the strong coupling limit
it is even possible to realize physical situations
in which two material quantum systems can be postselected in
a perfect Bell state by an appropriate von Neumann measurement of the quantized
radiation field. However, for this purpose
it is necessary to control 
the relevant interaction times between the quantized
radiation field and the two material few-level systems appropriately.
For sufficiently intense radiation fields these interaction times can even be chosen so short
that effects of spontaneous emission of photons into other modes of the radiation field can be neglected so that
a major decoherence mechanism can be eliminated and
all advantages of quantum interferences can be exploited.

In this paper we focus on the exploration of theoretical limits governing
entanglement creation between distant material qubits by 
postselective field measurements in the resonant quantum electrodynamical interaction regime.
Experimental realizations of the theoretical scenario discussed require 
two material qubits each of which is placed inside an optical cavity. The two cavities are connected by a quantum transmission
channel, such as an optical fiber,
which allows the transmission of the radiation field from one cavity to the other. Although the experimental realization of an
efficient coherent transfer of photons between cavities still constitutes a major technological challenge, methods for
coping with these challenges have been discussed previously \cite{Cirac,Enk,Pellizzari}.
In particular, recently developed sophisticated experimental
techniques \cite{Steinmetz,Colombe}
constitute important steps towards achieving almost perfect coherent couplings between a single mode of the
radiation field of a Fabry-P\'erot cavity and an optical fiber.

This paper is organized as follows.
In Sec.\ref{The model} we introduce the quantum 
electrodynamical model in which two elementary material
three-level systems interact 
with local cavity fields which are coupled by an optical fiber. 
Furthermore, we develop the general framework for 
describing the postselection procedure which prepares 
distant material qubits in a Bell state by 
an optimal generalized field measurement which introduces minimum errors. Numerical results are presented for characteristic
 quantities which quantify the
success with which a Bell pair is prepared, its fidelity,
and the minimum error with which this postselection can be achieved.
Whereas Sec.\ref{The model} discusses cases in which the propagation
of the optical radiation field between the two qubits 
through an optical fiber
is ideal,
in Sec.\ref{realistic} modifications
originating from photon loss during this propagation process
are taken into account.
In an Appendix we describe the photonic quantum
state transfer between two distant optical cavities connected
by a long optical fiber and we determine the conditions for
perfect photonic quantum state transfer.

\section{Photon-assisted entanglement creation}
\label{The model}
In this section we discuss a quantum electrodynamical model 
in which two spatially well separated elementary (material)
three-level systems are entangled in a Ramsey-type interaction scenario
with single-mode photonic quantum states
inside cavities connected by a long optical fiber. 
In particular, we explore the potential of producing
high-fidelity material Bell states with the help of  postselection by minimum-error
field measurements which are capable of distinguishing non orthogonal photonic states
optimally.
Thereby, we exploit
the fact that perfect
photonic quantum state transfer of single-mode quantum states
is possible between two optical cavities connected by an optical fiber
provided the cavity-fiber couplings are engineered appropriately (compare with Appendix A).
Thus, generalizing a recent proposal of van Loock et al. \cite{vanLoock1} to the almost resonant
strong
quantum electrodynamical coupling regime we demonstrate that this dynamical regime combined
with optimal postselection by field measurements offers interesting perspectives for producing
distant material Bell pairs with high fidelities and with high success probabilities.
\begin{figure}[b]
\includegraphics[width=7cm]{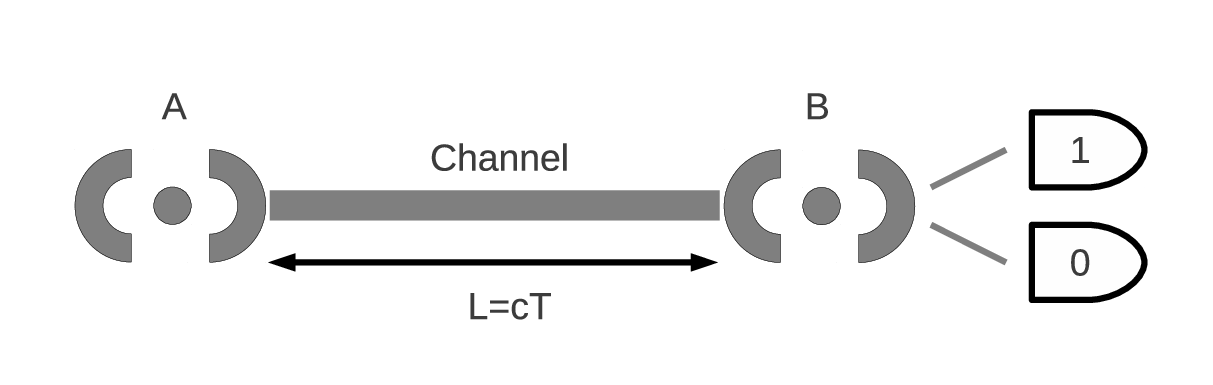}
\includegraphics[width=4.5cm]{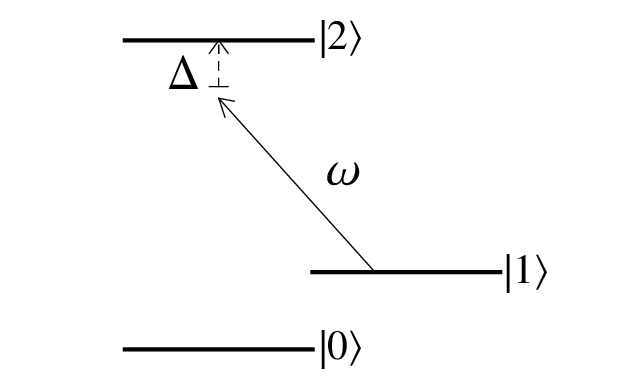}
\caption{\label{fig-1}Schematic representation of photon-assisted entanglement creation: 
The field state inside cavity $A$ interacts almost resonantly for a short time
$\tau$ with the material quantum system inside cavity $A$. The resulting photon state is 
transferred in a time $T = L/c\gg \tau$ to cavity $B$ by propagation through a connecting optical fiber.
($c$ is the propagation speed in the optical fiber.)
By appropriate engineering of the cavity-fiber couplings this quantum state transfer can be achieved perfectly.
After this quantum state transfer an analogous second almost resonant
short interaction
takes place for a time $\tau$. After this Ramsey-type interaction scenario the photon state
of cavity $B$
is measured 
by a minimum-error two-valued positive operator-valued measure
(POVM measurement) with measurement results $m\in \{1,0\}$.
The measurement result $m=1$ prepares both material quantum systems approximately
in a Bell state $|\Psi^+\rangle$ with success probability $P_{Bell}$ and with fidelity
$F_{opt}$.
}
\end{figure}

\subsection{The quantum electrodynamical model\label{QED}}

We consider two spatially separated optical cavities, say $A$ and $B$, which are connected by
a long optical fiber (compare with Fig. \ref{fig-1}). We assume that the coupling between these cavities and their connecting
optical fiber
is engineered in such a way that it is possible to transfer a quantum state prepared in a single
mode of cavity $A$, say  with frequency $\omega$,  perfectly to a single mode of cavity $B$ with the same
frequency. As demonstrated
in detail in the Appendix this is possible if these modes of cavities $A$ and $B$
are coupled resonantly to the densely spaced modes of the long connecting optical fiber and if the
coupling constants between cavities and optical fiber are engineered in such a way that a time reversal
of the decay process from cavity $A$ to the optical fiber is possible. In particular, this implies
that the decay rates of the relevant modes of
both cavities have to be equal, i.e. $\Gamma_A = \Gamma_B \equiv \Gamma$.
The total time $T$ required for such a
perfect photonic quantum state transfer between both cavities is determined by the length $L$ of the long
optical fiber and by the photonic propagation speed $c$ inside the fiber, i.e.
$T = L/c$. Furthermore, this time has to be large in comparison with the decay times of both cavities
so that both the leakage out of cavity $A$ into the fiber and from the fiber
into cavity $B$ can be completed, i.e.$T \gg 2/\Gamma$. 
Recent experimental advances constitute highly promising steps towards
realizing such links between cavities
and an optical fiber \cite{Steinmetz}.

In generalization of the original quantum repeater proposal of van Loock et al. \cite{vanLoock1}
we consider two elementary (material) three-level systems, such as trapped atoms or ions,
with energy eigenstates  $|0\rangle_i, |1\rangle_i$, and $|2\rangle_i$ ($i\in \{A,B\}$) and associated energies $E_0, E_1$, and
$E_2$ (compare with Fig.\ref{fig-1}). One of them ($i=A$) is located in cavity $A$ and the other one
($i=B$) in cavity $B$. 
The lowest energy eigenstates $|0\rangle_i$ and $|1\rangle_i$ 
are assumed to be hyperfine-split components
of the ground state with long
radiative lifetimes so that spontaneous decay of these states by photon emission can be neglected.
In the following these two states serve as the qubits which are going to be entangled.
The energy eigenstates 
$|1\rangle_i$ and $|2\rangle_i$ are assumed to be of opposite parity and to be
coupled almost resonantly to the single mode of frequency $\omega$ of cavity $i\in \{A,B\}$ by an optical
dipole transition.
The coupling of the far-detuned third level $|0\rangle_i$ 
to the radiation field is assumed to be negligible. 
It is the main purpose of our subsequent discussion 
to demonstrate the creation of entanglement between
the two qubits formed by the
states $|0\rangle_i$ and $|1\rangle_i$ ($i\in \{A,B\}$) 
with the help of almost resonant quantum electrodynamical couplings between
states $|1\rangle_i$ and
$|2\rangle_i$ ($i\in \{A,B\}$) and their respective local cavity fields which are correlated by photonic
quantum state transfer through the connecting optical fiber. 

For this purpose we consider a 
Ramsey-type interaction scenario which starts from
an initial state of the total matter-field system of the form
\begin{eqnarray}
\ket{\Psi(t=0)}&=&\frac{\ket{0}_A+\ket{1}_A}{\sqrt{2}}\otimes\frac{\ket{0}_B+\ket{1}_B}{\sqrt{2}} 
\otimes \ket{\alpha}_{A}\otimes\ket{0}.\nonumber\\
\label{initialstate} 
\end{eqnarray}
Thus, the two spatially well separated material qubits are prepared in a particular
separable state and
the single mode of frequency $\omega$ inside cavity $A$ is prepared in a coherent state $|\alpha\rangle$.
All other field modes of the radiation field involved are prepared in the vacuum state.
After this preparation in a first step the single-mode radiation field inside cavity $A$
interacts with the three-level system $A$
almost resonantly for a time 
$\tau$ which is assumed to be small in comparison with the decay time of this cavity mode, 
i.e. $\tau \ll 1/\Gamma$.
We also assume that this interaction time is so small that spontaneous emission of photons
from the excited state $|2\rangle_A$ is negligible. 
Such a short and almost resonant interaction between the material three-level system 
and the field mode inside cavity $A$ can be turned on and turned off
by Stark switching techniques, for example, by which 
the dipole transition between levels $|1\rangle_A$ and $|2\rangle_A$ is tuned 
with the help of an externally applied
electric field. After this almost instantaneous
interaction (on the time scale of the cavity decay) the resulting photonic quantum state inside cavity $A$
is strongly entangled with 
the three-level system $A$ and 
propagates to cavity $B$ through the optical fiber. At time $T = L/c \gg 2/\Gamma \gg \tau$
the photonic quantum state produced by the almost instantaneous interaction inside cavity $A$
has been transferred to the single mode of frequency $\omega$ in cavity $B$.
In the second step of the Ramsey-type interaction scenario at time $T$ an analogous second almost resonant
interaction of the photon field
with the second three-level system is turned on and off for the short time 
$\tau \ll 2/\Gamma \ll T$ inside cavity $B$. 
After these two almost instantaneous
matter-field interactions the resulting photon state is measured by photon detectors.

In our subsequent discussion we are interested in cases
in which the interaction times $\tau$ in cavities $A$ and $B$ are
significantly shorter than the decay time of the cavities and
the radiative life time of the excited
states $|2\rangle_i$ ($i\in \{A,B\}$) so that effects of spontaneous decay from these levels during the interaction time
can be neglected. In order to ensure that effects of spontaneous decay from these excited states 
$|2\rangle_i$ ($i\in \{A,B\}$)
are negligible also during the long propagation of the photons from cavity $A$ to cavity $B$
one may transfer the excitations of these 
levels coherently to radiatively stable hyperfine split ground state components, 
say $|\tilde{2}\rangle_i$ ($i\in \{A,B\}$), 
with the help of two (possibly  classical) $\pi$ pulses applied immediately after the
interaction of each three-level system with
its local single-mode photon field. 
Thus, replacing in our subsequent theoretical considerations 
the excited states
$|2\rangle_i$ ($i\in \{A,B\}$) and their energy $E_2$ by their corresponding radiatively stable 
states $|\tilde{2}\rangle_i$ ($i\in \{A,B\}$) with corresponding energy $\tilde{E}_2$
effects of spontaneous emission can be neglected during all stages of this interaction scenario.

The dynamics of the short almost resonant interaction between the three-level system $j\in \{A,B\}$
and the occupied
local field mode inside cavity $j$ is described by the Hamiltonian
\begin{eqnarray}
\hat{H}_j&=& \hat{H}_j^{(Q)} + \hbar \omega \hat{a}_j^{\dagger}\hat{a}_j +
\hbar g \hat{a}_j|2\rangle_{jj}\langle 1|+\hbar g^* \hat{a}^{\dagger}_j|1\rangle_{j}\langle 2|,\nonumber\\
\label{Hamiltonian10}
\end{eqnarray}
with the unperturbed  
Hamiltonian $\hat{H}_j^{(Q)} = \sum_{k=0,1,2} E_k |k\rangle_{jj}\langle k|$
of the three-level system $j$ 
.
Thereby,
the interaction between the local optical field modes and the material systems
$A$ and $B$ is described in the dipole and rotating wave approximation. It is
characterized by the coupling parameters 
$g_j = -~_j\langle 2|\hat{{\bf d}}|1\rangle_j\sqrt{\hbar \omega/(2\epsilon_0)}{\bf g}_j({\bf x}_j)/\hbar$ ($j\in \{A,B\}$) which involve
the transition dipole moments $~_j\langle 2|\hat{\bf d}|1\rangle_j$ and the normalized mode functions of the
single mode radiation field ${\bf g}_j({\bf x})$ in their respective cavities. 
In the following 
we concentrate on cases with symmetric couplings, i.e. $g = g_A = g_B $. The modulus of this
characteristic coupling strength defines the resonant vacuum Rabi frequency $\Omega_{vac} 
~=~ \mid g\mid $ 
\cite{Schleich}. 
The photonic annihilation (creation) operators of the cavity-modes $A$ and $B$ are 
denoted by
$\hat{a}_A$ and $\hat{a}_B$ ($\hat{a}^\dagger_A$ and $\hat{a}^\dagger_B$). 
with the corresponding Fock states  $\ket{n}_{A}$ and $\ket{n}_{B}$. 

It is straightforward to determine the quantum state 
of the two cavity modes and the two three-level systems
at time $t=T$ after the Ramsey-type interaction sequence
described above. If we assume that 
perfect quantum state 
transfer is achieved between the two cavities coupled by
the optical fiber(see Appendix \ref{Appendix}) the pure state $|\Psi(t)\rangle$ of both
material three-level systems and of the field mode inside cavity $B$ is given by
\begin{eqnarray}
\label{Psi}
&&\ket{\Psi(t)} =
|g_5(t)\rangle |1\rangle_A |2\rangle_B \ee^{-i\Phi_{12}} + 
|g_6(t)\rangle 
|2\rangle_A |2\rangle_B \ee^{-i\Phi_{22}}  + \nonumber
\\
&&\frac{1}{2}|0\rangle_A |0\rangle_B \ket{\alpha\ee^{-i\omega t}}\ee^{-i\Phi_{00}}+
|g_1(t)\rangle |\Psi^{+}\rangle_{AB}\ee^{-i\Phi_{10}} +\nonumber\\ 
&&|g_2(t)\rangle \left(|0\rangle_A |2\rangle_B \ee^{-i\Phi_{02}} + 
|2\rangle_A |0\rangle_B \ee^{-i\Phi_{20}}\right)
+\nonumber\\
&&|g_3(t)\rangle |1\rangle_A |1\rangle_B \ee^{-i\Phi_{11}}
+
|g_4(t)\rangle |2\rangle_A |1\rangle_B \ee^{-i\Phi_{21}}
\end{eqnarray}
with the Bell state $|\Psi^+\rangle_{AB} = (|0\rangle_A |1\rangle_B + |1\rangle_A |0\rangle_B)/\sqrt{2}$ involving
the two distant qubits of systems $A$ and $B$. 
The time-dependent phases
$\Phi_{00} = 2E_0 t/\hbar,
\Phi_{10}=(E_0  + E_1)t/\hbar + \Delta \tau/2,
\Phi_{20}=(E_0  + E_2)t/\hbar  - \Delta\tau/2, 
\Phi_{02}=(E_0 + E_1 + \hbar \omega)t/\hbar +\Delta\tau/2,
\Phi_{11}=2E_1t/\hbar + \Delta \tau,
\Phi_{12}= (2E_1 + \hbar \omega)t/\hbar + \Delta \tau,
\Phi_{21}= (E_1 + E_2)t/\hbar,
\Phi_{22}=(E_1 + E_2 + \hbar \omega)t/\hbar$ 
characterize
the  interferences appearing in
this Ramsey-type interaction scenario. The detuning from resonance is denoted by
$\Delta := (E_2-E_1)/\hbar - \omega$.
The (unnormalized) pure field states $|g_j(t)\rangle$ ($j=1,...,6$) and $\ket{\alpha\ee^{-i\omega t}}$
entering Eq.(\ref{Psi})
are defined by 
\begin{widetext}
\begin{eqnarray}
\label{noassump}
\ket{\alpha\ee^{-i\omega t}}&=&\sum_{n=0}^{\infty}
\ee^{-|\alpha|^2/2}\frac{\alpha^n \ee^{-in\omega t}}{\sqrt{n!}}|n\rangle_{B},\nonumber \\
 |g_1(t)\rangle &=&
\sum_{n=0}^{\infty}
\ee^{-|\alpha|^2/2}\frac{\alpha^n}{\sqrt{n!}}
\frac{1}{\sqrt{2}}
\Big(\cos\big(\Omega(n)\tau\big)+i\frac{\Delta}{2\Omega(n)} \sin\big(
\Omega(n)\tau\big) \Big) \ee^{-i \omega n t}|n\rangle_{B}, \nonumber \\
 |g_2(t)\rangle &=&-\sum_{n=0}^{\infty}
\ee^{-|\alpha|^2/2}\frac{\alpha^{(n+1)}}{\sqrt{(n+1)!}}\frac{ig\sqrt{n+1}}
{\Omega(n+1)} \frac{1}{{2}}
\sin\big( \Omega(n+1) \tau\big)\ee^{-i \omega n t}|n\rangle_{B}, \nonumber \\
|g_3(t)\rangle &=&
\sum_{n=0}^{\infty}\ee^{-|\alpha|^2/2}\frac{\alpha^n}{\sqrt{n!}}\frac{1}{2}\Big(\cos\big(\Omega(n)\tau\big)+i
\frac{\Delta}{2\Omega(n)} \sin\big(
\Omega(n)\tau\big) \Big)^2\ee^{-i \omega n t}|n\rangle_{B}, \nonumber \\
|g_4(t)\rangle &=&-\sum_{n=0}^{\infty}\ee^{-|\alpha|^2/2}\frac{\alpha^{(n+1)}}{\sqrt{(n+1)!}}
\frac{ig\sqrt{n+1}}{\Omega(n+1)} \sin\big( \Omega(n+1) \tau\big)\frac{1}{2}
\Big(\cos\big(\Omega(n)\tau\big)+i\frac{\Delta}{2\Omega(n)} \sin\big(
\Omega(n)\tau\big) \Big)\ee^{-i \omega n t}|n\rangle_{B}, \nonumber \\
|g_5(t)\rangle &=& -\sum_{n=0}^{\infty}\ee^{-|\alpha|^2/2}\frac{\alpha^{(n+1)}}{\sqrt{(n+1)!}}
\frac{ig\sqrt{n+1}}{\Omega(n+1)} \frac{\sin( \Omega(n+1) \tau)}{2}
\Big(\cos\big(\Omega(n+1)\tau\big)+i\frac{\Delta}{2\Omega(n+1)} \sin\big(\Omega(n+1)\tau\big) \Big)\ee^{-i 
\omega n t}|n\rangle_{B}, 
\nonumber \\
|g_6(t)\rangle &=&-\sum_{n=0}^{\infty}\ee^{-|\alpha|^2/2}\frac{\alpha^{(n+2)}}{\sqrt{(n+2)!}}\
\frac{g\sqrt{n+1}}{\Omega(n+1)} \sin\big( \Omega(n+1) \tau\big)
\frac{g\sqrt{n+2}}{\Omega(n+2)}\frac{1}{2} \sin\big( \Omega(n+2) \tau\big)\ee^{-i \omega n t}|n\rangle_{B}
\end{eqnarray}
\end{widetext}
with the normalized photon number states $|n\rangle_{B}$ ($n\in {\mathbb N}_0$). The parameter 
$\Omega(n) := \sqrt{\Delta^2/4 + \mid g\mid^2n}$ denotes the effective Rabi frequency associated with 
the photon number $n$ of the optical radiation field in cavity $B$. 

The quantum state of Eq.(\ref{Psi}) yields a complete description of the interaction between the material
quantum systems $A$ and $B$
and the optical radiation 
fields in the idealized case of perfect quantum state transfer between cavities $A$ and $B$
mediated by propagation
through the connecting optical fiber.
In particular, it clearly exhibits the resulting entanglement
between the material systems $A$ and $B$ on the one hand and the radiation field inside cavity $B$
on the other hand.
In the weak coupling limit of large detunings from resonance this latter entanglement has been used in the proposal
by van Loock et al. \cite{vanLoock1} for creating an entangled Bell state $|\Psi^{+}\rangle_{AB}$
by projecting out the field state $|g_1(t)\rangle$ by
a generalized positive-operator-valued quantum measurement (POVM) \cite{Helstrom,Holevo,Hayashy} performed on the optical radiation field.
However, in the weak coupling limit 
the field states $|g_i(t)\rangle$ ($i=1,...,6$)
are not orthogonal so that the field state
$|g_1(t)\rangle$ cannot be distinguished perfectly from the other field states.
This limits the
achievable entanglement of the entangled state of the two material qubits
significantly. In our subsequent subsection it will be demonstrated that in the strong coupling limit of almost resonant 
quantum electrodynamical coupling these limitations can be circumvented and in certain dynamical regimes even perfect
Bell states $|\Psi^+\rangle_{AB}$ can be prepared by suitable quantum measurements performed
on the optical field inside cavity $B$.

\subsection{Postselection of Bell states by minimum-error field measurements}

In this section we investigate to what extent the Ramsey-type interaction scenario discussed in the previous section
can be optimized in order to prepare a maximally entangled Bell state between the distant
material quantum systems $A$ and $B$
by an appropriate minimum-error POVM measurement of the optical field in cavity $B$.
This aspect is of particular interest for future realizations of 
quantum repeaters which require high-fidelity Bell pairs as a resource.
It is demonstrated that in the ideal case of perfect state transfer of
the optical radiation field through 
the fiber from cavity
$A$ to system $B$
in the strong coupling limit of almost resonant
interaction between the local cavity fields and the material three-level systems
high-fidelity Bell states of the material systems $A$ and $B$ can be prepared 
provided the relevant interaction times $\tau$
are controlled appropriately. 

Let us start from the pure quantum state $|\Psi(t)\rangle$ of Eq.(\ref{Psi}) 
which describes the entanglement of the matter-field system
after the Ramsey-type interaction scenario with the single occupied field mode inside cavity $B$. 
The resulting reduced density operator of the field state $\hat{\rho}_F(t)$
in cavity $B$ is 
obtained by tracing out the material degrees of freedom, i.e.
\begin{eqnarray}
\hat{\rho}_F(t) &=& {\rm Tr}_{AB}\{ |\Psi(t)\rangle \langle \Psi(t)| \} =   p \hat{\rho}_1 + (1-p) \hat{\rho}_2  
\label{fieldstate1}
\end{eqnarray}
with the field states
\begin{eqnarray}
\hat{\rho}_1 &=&\frac{|g_1(t)\rangle \langle g_1(t)|}{p},\nonumber\\
\hat{\rho}_2 &=&\Big(\frac{1}{4}|\alpha e^{-i\omega t}\rangle \langle \alpha e^{-i\omega t}| 
+ 2|g_2(t)\rangle\langle g_2(t)|\nonumber\\
&+& \sum_{j=3}^{6}|g_j(t)\rangle \langle g_j(t)|\Big)/(1-p)\nonumber\\
&&
\label{fieldstate}
\end{eqnarray}
and with $p =\langle g_1(t) |g_1(t)\rangle$
denoting the a priori probability of the pure field state $|g_1(t)\rangle$.
In general, the quantum states $\hat{\rho}_1$ and $\hat{\rho}_2$ are not orthogonal 
so that they cannot be distinguished by any quantum measurement perfectly \cite{Chefles,Chefles2,Bruss}.

Therefore, in order to optimize the postselection of a perfectly 
entangled Bell state $|\Psi^+\rangle_{AB}$ it is necessary to perform a POVM measurement on the 
optical radiation field with two possible measurement results $m$, say $m\in \{1,0\}$. The first measurement result
$m=1$ indicates an optimal projection of the field state $\hat{\rho}_F(t)$
onto the pure field state $\hat{\rho}_1$
and the second measurement result $m=0$,
indicates an optimal projection of $\hat{\rho}_F(t)$ onto the mixed field state  $\hat{\rho}_2$.
Let us denote the positive operators  associated with these two measurement results by
$\hat{T}_1\geq 0 $ and $0 \leq \hat{T}_0 := {\bf I} - \hat{T}_1$.
(${\bf I}$ denotes the unit operator in the Hilbert space of the single-mode
radiation field.) The positive operator $\hat{T}_1$ of this POVM $\{\hat{T}_1, \hat{T}_0\}$
has to be determined in such
a way that for a given a priori probability $p$ the error probability
\begin{eqnarray}
E &=& p {\rm Tr}\{\hat{\rho}_1\hat{T}_0\} + (1-p){\rm Tr}\{\hat{\rho}_2\hat{T}_1\} 
\end{eqnarray}
is as small as possible. Diagonalizing the Hermitian operator $\hat{A}:= p\hat{\rho}_1 - (1-p)\hat{\rho}_2 $ according to
$\hat{A} = \sum_{\lambda} \lambda |\lambda\rangle\langle \lambda|$
the solution of this optimization problem is given by the projection operator \cite{Helstrom,Holevo,Hayashy}
\begin{eqnarray}
\hat{T}_1 &=& 
\sum_{\lambda\geq 0} |\lambda\rangle\langle \lambda| = {\bf I} - \hat{T}_0
\label{optimalPOVM}
\end{eqnarray}
which
projects onto the non-negative spectral components of the operator $\hat{A}$.
The corresponding minimal error probability $E_{min}$ of the optimal POVM measurement defined by
Eq.(\ref{optimalPOVM}) 
is determined by the trace norm distance between the two (unnormalized) components 
$p\hat{\rho}_1$ and $(1-p)\hat{\rho}_2$
of the quantum state $\hat{\rho}_F(t)$, i.e. \cite{Helstrom,Hayashy}
\begin{eqnarray}
E_{min} = \frac{1}{2}\left( 1 - ||p \hat{\rho}_1 - (1-p) \hat{\rho}_2||_1 \right).
\label{EMin}
\end{eqnarray}
The probability $P_{Bell}$ with which this optimal POVM measurement of the optical radiation field
prepares the distant material
quantum systems $A$ and $B$ in the Bell state $|\Psi^+\rangle$ it thus given by
\begin{eqnarray}
P_{Bell} = p{\rm Tr}_{field}\{\hat{\rho}_1 \hat{T}_1\}.
\label{PBell}
\end{eqnarray}

>From Eqs.(\ref{EMin}) and (\ref{PBell}) it is apparent that
if the quantum states $\hat{\rho}_1$ and $\hat{\rho}_2$ were orthogonal the positive operator $\hat{T}_1$ of the POVM measurement
would be
a projection operator onto the support of the state $\hat{\rho}_1$ and the success probability $P_{Bell}$ 
would be given by $p$. In addition, the minimal error probability $E_{min}$ would vanish. However, the
typical non-orthogonality of the field states $\hat{\rho}_1$ and $\hat{\rho}_2$ complicates matters and causes unavoidable errors
even if minimum-error POVM measurements are performed.

With the optimal POVM measurement $\{\hat{T}_1,\hat{T}_0\}$ as defined by Eq.(\ref{optimalPOVM})
we can associate a deterministic quantum operation \cite{NielsenChuang} with the Kraus operators
$\{\hat{U}_1\sqrt{\hat{T}_1}, \hat{U}_0\sqrt{{\bf I} - \hat{T}_1}\}$ which allows us to determine the quantum state of the matter-field
system associated with the two measurement results $m\in \{1,0\}$ of the optimal POVM. 
Thereby, the linear operators $\hat{U}_1$ and $\hat{U}_0$
are partial isometries between the ranges of  $\sqrt{\hat{T}_1}$ and $\sqrt{\hat{T}_0}$ and the Hilbert space of the optical radiation
field. 
After a successful
POVM measurement
the state of both material quantum systems $A$ and $B$ 
is given by
\begin{eqnarray}
\hat{\rho}_{AB}(t) &=&
\frac{
{\rm Tr}_{fields}
\{
|\Psi(t) \rangle \langle \Psi(t)|
\hat{T}_1
\}
}{
{\rm Tr}_{A,B,fields}
\{
|\Psi \rangle \langle \Psi|
\hat{T}_1
\}
}. 
\end{eqnarray}
Note that this quantum state of systems $A$ and $B$ is independent of the transformation $\hat{U}_1$.
Thus,
the fidelity $F_{opt}$ of an optimally prepared Bell pair which is postselected by a measurement result with value $m=1$ 
(corresponding to the POVM operator $\hat{T}_1$) is given by
\begin{eqnarray}
F_{opt} &=&
\sqrt{\langle \Psi^+| \hat{\rho}_{AB}(t) |\Psi^+\rangle}.
\label{OptFid}
\end{eqnarray}

In order to obtain some insight into the dynamical parameter regimes in which 
this postselection procedure
may yield
high-fidelity Bell pairs 
let us concentrate on the practically important case
of large mean photon numbers, i.e. $\overline{n} = |\alpha|^2 \gg 1$, and on intermediate values of the interaction times $\tau$ of the
optical radiation field with the quantum systems $A$ and $B$ so that in Eq.(\ref{Psi})
the photon-number dependent Rabi frequencies $\Omega(n)$ can
be linearized around the mean photon number $\overline{n}$. This linearization is valid if the condition
\begin{eqnarray}
&&
\frac{1}{2}\mid \frac{d^2\Omega}{dn^2}(n)\mid_{n=\overline{n}}\tau  (\Delta n)^2 =
\frac{1}{8\overline{n}}\frac{|\overline{\Omega}_0|^4 \tau}{|\overline{\Omega}|^3} \ll \pi
\label{intermediate}
\end{eqnarray}
is fulfilled 
so that revival phenomena \cite{Schleich} can be neglected. 
Thereby, $\Delta n = |\alpha| = \sqrt{\overline{n}}$ denotes the photon number uncertainty of the coherent
state $|\alpha \rangle$ and $\overline{\Omega} = \sqrt{\Delta^2/4 + |g|^2 \overline{n}}$
and $\overline{\Omega}_0 = |g|\sqrt{\overline{n}}$ are the effective Rabi frequency and the resonant Rabi frequency associated with
the mean photon number $\overline{n}$. 
In this linearization the field states of Eq.(\ref{Psi}) can be approximated by
\begin{widetext}
\begin{eqnarray}
\label{coherent}
\ket{g_1} &=&
\ket{\alpha \ee^{-i\omega t}\ee^{i \theta}}
e^{i\omega_c \tau}
\frac{1+\Delta/(2\overline{\Omega})}{2\sqrt{2}}
+\ket{\alpha \ee^{-i\omega t} \ee^{-i \theta}}
e^{-i\omega_c \tau}
\frac{1-\Delta/(2\overline{\Omega})}{2\sqrt{2}},\nonumber\\
\ket{g_2}&=&-e^{i\varphi}
\frac{\overline{\Omega}_0}{4\overline{\Omega}}\left( 
\ket{\alpha \ee^{-i\omega t}\ee^{i \theta}}
e^{i\omega_c \tau}
-\ket{\alpha \ee^{-i\omega t} \ee^{-i \theta}}
e^{-i\omega_c \tau}
\right),\nonumber\\
\ket{g_3}&=&
\ket{\alpha \ee^{-i\omega t}\ee^{2i \theta}}
e^{2i\omega_c \tau}\frac{(1+{\Delta}/(2\overline{\Omega}))^2}{8} +
\ket{\alpha \ee^{-i\omega t}\ee^{-2i \theta}}
e^{-2i\omega_c \tau}\frac{(1-{\Delta}/(2\overline{\Omega}))^2}{8} +
\ket{\alpha \ee^{-i\omega t}}
\frac{1-({\Delta}/(2\overline{\Omega}))^2}{4},\nonumber\\ 
\ket{g_4}
&=&
\ket{g_5} =
-e^{i\varphi}\frac{\overline{\Omega}_0}{8\overline{\Omega}}\left(
\ket{\alpha \ee^{-i\omega t}\ee^{2i \theta}}
e^{2i\omega_c \tau}(1+\frac{\Delta}{2\overline{\Omega}}) -
\ket{\alpha \ee^{-i\omega t}\ee^{-2i \theta}}
e^{-2i\omega_c \tau}(1-\frac{\Delta}{2\overline{\Omega}}) -
\ket{\alpha \ee^{-i\omega t}}\frac{\Delta}{\overline{\Omega}}
\right),\nonumber\\
\ket{g_6}
&=&e^{2i\varphi}
\frac{\overline{\Omega}^2_0}{8\overline{\Omega}^2}\left(
\ket{\alpha \ee^{-i\omega t}\ee^{2i \theta}}
e^{2i\omega_c \tau} +
\ket{\alpha \ee^{-i\omega t}\ee^{-2i \theta}}
e^{-2i\omega_c \tau} -
2\ket{\alpha \ee^{-i\omega t}}
\right)
\end{eqnarray}
\end{widetext}
with $g = e^{i\varphi}|g|$, 
$\theta=\overline{\Omega}^2_0 \tau/(2\overline{\Omega} \overline{n}) \ll 1$, and
with the modified effective Rabi frequency $\omega_c= \overline{\Omega}[1 - \overline{\Omega}^2_0/(2\overline{\Omega}^2)]$.
This linearization implies that the a priori probability $p$ entering Eq.(\ref{fieldstate1}) can be approximated by 
\begin{eqnarray}
p  &=&
\frac{\Delta^2}{8\overline{\Omega}^2} + 
\left( \frac{1}{4} - \left(\frac{\Delta}{4\overline{\Omega}}\right)^2 \right)\times\nonumber\\
&&\left(1 +
\cos(2\overline{\Omega}\tau)\exp(-(\overline{\Omega}^2_0\tau/(\overline{\Omega}\sqrt{\overline{n}}))^2/2)
\right). 
\label{apriori}
\end{eqnarray}
Furthermore,
the overlaps between the coherent state
$|g_1(t)\rangle$ and the states
$|\alpha e^{i\omega t}\rangle$ and $|g_3(t)\rangle$
constituting significant parts of the quantum state
$\hat{\rho}_2$ 
reduce to
\begin{widetext}
\begin{eqnarray}
\langle \alpha e^{-i\omega t}|g_1(t)\rangle &=&
\frac{1}{\sqrt{2}}
\exp(-(\overline{\Omega}^2_0\tau/(\overline{\Omega}\sqrt{\overline{n}}))^2/2)
\left(\cos(\overline{\Omega}\tau)
 + i\frac{\Delta}{2\overline{\Omega}}\sin(\overline{\Omega}\tau))
\right),\nonumber\\
\langle g_3(t)| g_1(t)\rangle &=& 
e^{-[\overline{\Omega}_0^2\tau/(2\overline{\Omega}\sqrt{\overline{n}})]^2/2}
\left(
e^{-i\overline{\Omega}\tau}\frac{(1+\Delta/(2\overline{\Omega}))^3}{16\sqrt{2}} +
e^{i\overline{\Omega}\tau}\frac{(1-\Delta/(2\overline{\Omega}))^3}{16\sqrt{2}} 
\right) + \nonumber\\
&&
e^{-9[\overline{\Omega}_0^2\tau/(2\overline{\Omega}\sqrt{\overline{n}})]^2/2}
\left(
e^{-3i\overline{\Omega}\tau}\frac{(1+\Delta/(2\overline{\Omega}))^2 (1-\Delta/(2\overline{\Omega}))}{16\sqrt{2}} +
e^{3i\overline{\Omega}\tau}\frac{(1+\Delta/(2\overline{\Omega})) (1-\Delta/(2\overline{\Omega}))^2}{16\sqrt{2}} 
\right) + \nonumber\\
&&
e^{-[\overline{\Omega}_0^2\tau/(2\overline{\Omega}\sqrt{\overline{n}})]^2/2}
\left(
e^{i\overline{\Omega}\tau}\frac{(1+\Delta/(2\overline{\Omega})) [1-(\Delta/(2\overline{\Omega}))^2]}{8\sqrt{2}} +
e^{-i\overline{\Omega}\tau}\frac{(1-\Delta/(2\overline{\Omega})) [1-(\Delta/(2\overline{\Omega}))^2]}{8\sqrt{2}} 
\right).  
\label{overlap10}
\end{eqnarray}
\end{widetext}

>From Eq.(\ref{overlap10}) it is apparent that in the case of resonant excitation, i.e. $\Delta =0$,
these overlaps vanish at interaction times
$\tau = (\pi + 2k \pi)/(2\overline{\Omega}_0)$ ($k \in {\mathbb N}_0$). 
Furthermore, at these particular interaction times also all other overlaps $\langle g_j(t) |g_1(t)\rangle $ with $j=2,4,5,6$
vanish. Thus, in this linearization approximation
the quantum states $\hat{\rho}_1$ and $\hat{\rho}_2$ are orthogonal at these interaction times so that
they can be distinguished perfectly by a von Neumann measurement described by the projection operators
$\{\hat{T}_1 = \hat{\rho}_1, \hat{T}_0 = {\bf I} - \hat{T}_1\}$.
In this case the
success probability 
reduces to $p =  (1 - \exp(-(\overline{\Omega}_0\tau/\sqrt{\overline{n}})^2/2))/4$ and approaches
the value of $p =1/4$ in the limit of sufficiently large interaction times of the order of the inverse vacuum Rabi frequency, i.e. 
$\tau
\gg \sqrt{\overline{n}}/\overline{\Omega}_0 = 1/\Omega_{vac}$.
In addition, the error probability $E_{min}$ of this optimal von Neumann measurement vanishes
and the fidelity  $F_{opt}$ of the prepared Bell state $|\Psi^+\rangle$  equals unity.

This resonant behavior at these particular interaction times is in marked contrast to the behavior at large detunings
$ |\Delta/2| \gg |\overline{\Omega}_0|$ at which we obtain $\overline{\Omega} \to |\Delta/2|$ so that
the above mentioned overlaps tend to the non vanishing values
$|\langle \alpha e^{-i\omega t}|g_1\rangle| =
{\rm exp}\{-(\overline{\Omega}^2_0\tau/(\overline{\Omega}\sqrt{\overline{n}}))^2/2\}/\sqrt{2}$
and
$|\langle g_3| g_1\rangle| = 
{\rm exp}\{-[\overline{\Omega}_0^2\tau/(2\overline{\Omega}\sqrt{\overline{n}})]^2/2\}/(8\sqrt{2})$.
As a result, in this weak coupling limit it is only for extremely large
interaction times,
i.e. $\tau \geq |\Delta/\overline{\Omega}_0|/\Omega_{vac} \gg 1/\Omega_{vac}$, that these overlaps become small.
However, at these interaction times, which are significantly larger than the inverse
vacuum Rabi frequency, typically additional effects originating
from spontaneous emission of photons also
have to be taken into account which have been neglected in our analysis.
Thus, apart from these extremely large interaction times in the weak coupling limit it is 
never possible to distinguish the relevant
field states $\hat{\rho}_1$ and $\hat{\rho}_2$ perfectly so that the preparation of perfect Bell states $|\Psi^+\rangle$
is impossible.

Physically speaking, these marked differences between the strong and the weak coupling cases are due to the characteristic
dephasing phenomena 
which are also responsible for the well known collapse 
phenomena in the Jaynes-Cummings-Paul model \cite{Schleich}.
For the case of interaction times $\tau$ characterized by the inequalities (\ref{intermediate})
these interference phenomena are captured quantitatively by the various slightly shifted coherent states entering Eq.(\ref{overlap10}).
Although these coherent states of the form $|\alpha e^{-i\omega t}e^{i\theta k}\rangle$ with $\pm k = 0,1,2$ are shifted in their phases
only slightly by multiples
of the small amount $\theta = \overline{\Omega}_0^2 \tau/(2\overline{\Omega}\overline{n}) \ll 1$ their overlaps are given by
\begin{eqnarray}
&&|\langle \alpha e^{-i\omega t} e^{-ik\theta}|\alpha e^{-i\omega t} e^{-ik'\theta}\rangle| =\nonumber\\
&&
{\rm exp}\{- [(k'-k)\overline{\Omega}_0^2\tau/(\overline{\Omega}\sqrt{\overline{n}})]^2/2\}.
\end{eqnarray}
This implies that for very small interaction times $\tau$ or large mean photon numbers $\overline{n}$, i.e.
$\overline{\Omega}_0^2 \tau/(\overline{\Omega}\sqrt{\overline{n}})/2 \ll  1$,
the coherent states entering Eq.(\ref{overlap10}) 
are almost identical. 
Thus, in this limit the dynamics is well approximated by the semiclassical limit, i.e. $\overline{n}\to \infty$,
in which the influence of the single-mode radiation field can be approximated well by a classical field and the influence of
photon fluctuations is negligible. 
However, for intermediate interaction times of the order of 
$\overline{\Omega}_0^2 \tau/(\overline{\Omega}\sqrt{\overline{n}})/2 \geq  1$ the overlaps of the coherent states
entering Eq.(\ref{overlap10}) 
become small so that their interferences
are suppressed significantly.
This suppression of interference due to dephasing of these coherent states is also the reason for the appearance of the well known
collapse phenomena in the Jaynes-Cummings-Paul model. 
In particular, in the case of resonant excitation it implies that even for arbitrary interaction times of the order of
$\tau \geq \sqrt{\overline{n}}/\overline{\Omega}_0 := 1/\Omega_{vac}$ 
the field states
$\hat{\rho}_1$ and $\hat{\rho}_2$ are almost orthogonal so that they can be distinguished almost perfectly by an appropriate 
quantum measurement. Effects of spontaneous emissions of photons from the excited state $|2\rangle$ with
a spontaneous decay rate $\Gamma$ can still be neglected as long
as the matter-field coupling as characterized by the vacuum Rabi frequency
is sufficiently large so that $\Omega_{vac} \gg \Gamma$.
Meeting these requirements is within reach of current experimental possibilities.
The recent experiment by McKeever et al. \cite{Kimble}, for example, was performed in the optical frequency regime
and was characterized by the parameters $\Omega_{vac}/(2\pi) = 16 {\rm Mhz}$  and $\Gamma/(2\pi) = 2.6 {\rm Mhz}$.
More recent experiments have even reported vacuum Rabi frequencies
$\Omega_{vac}/(2\pi)$ exceeding $20$ gigahertz \cite{Colombe}.
However, in cases of large detunings, i.e. $|\Delta/2| \gg \overline{\Omega}_0$,
according to Eq.(\ref{overlap10}) strong dephasing requires significantly larger interaction times of the order of
$\tau \geq |\Delta/\overline{\Omega}_0|/\Omega_{vac} \gg 1/\Omega_{vac}$ for which typically the 
influence of spontaneous decay processes can no longer be neglected. 

\begin{figure}[t]
\begin{center}
\includegraphics[width=8.cm]{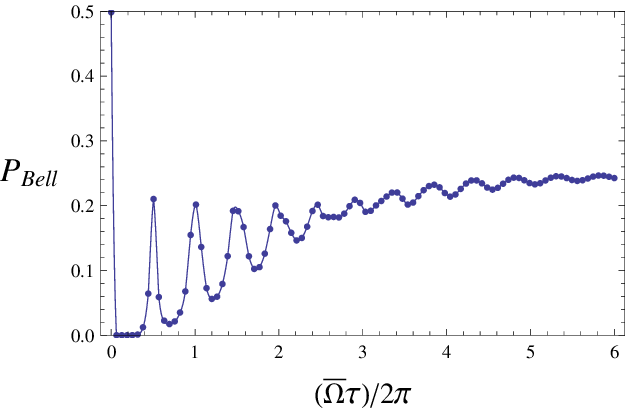}
\includegraphics[width=8cm]{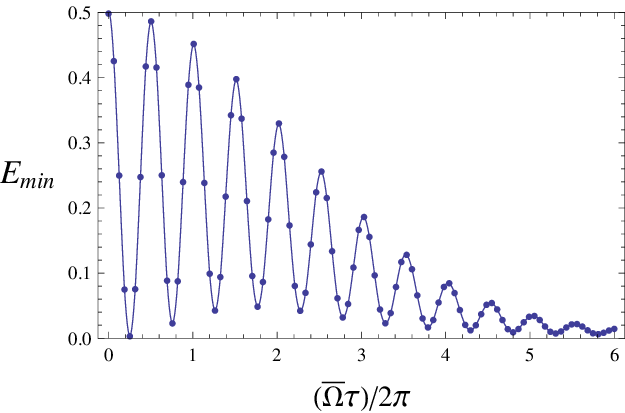}
\includegraphics[width=8cm]{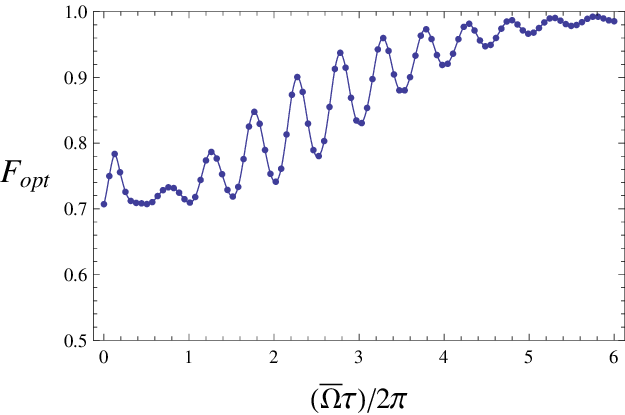}
\caption{\label{Fig1}Dependence of characteristic quantities on the interaction time $\tau$ in
the strong coupling limit $\Delta =0$: The success probability for postselecting a Bell state
$|\Psi^+\rangle_{AB}$ of
Eq.(\ref{PBell}) (top);
the minimum-error probability of the POVM measurement of the field of Eq.(\ref{EMin}) (middle);
the fidelity of the postselected
Bell state $|\Psi^+\rangle_{AB}$ of Eq.(\ref{OptFid}) (bottom).}
\end{center}
\end{figure}

\begin{figure}[t]
\begin{center}
\includegraphics[width=8.cm]{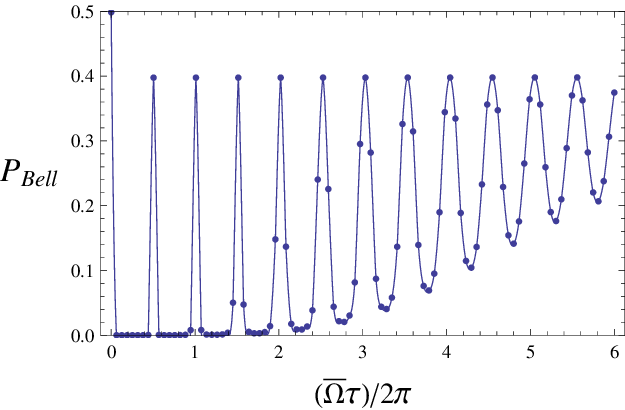}
\includegraphics[width=8cm]{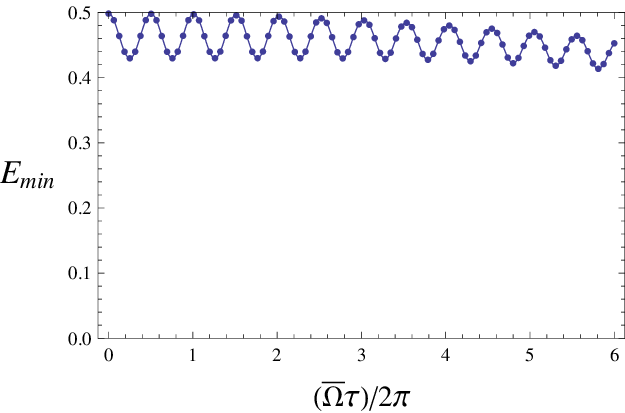}
\includegraphics[width=8cm]{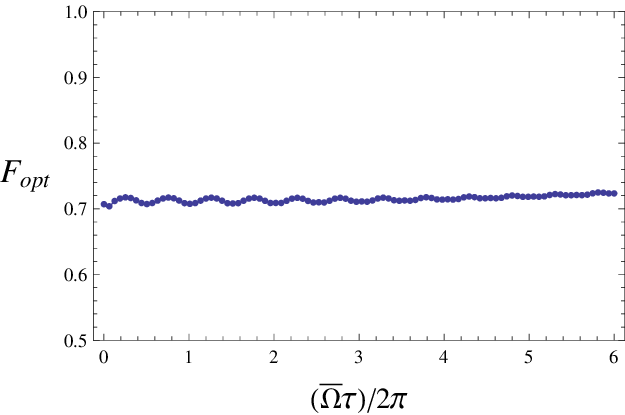}
\caption{\label{Fig2}Dependence of characteristic quantities on the interaction time $\tau$ in
the weak coupling limit $\Delta = 5 \overline{\Omega}_0$:
The success probability for postselecting a Bell state
$|\Psi^+\rangle_{AB}$ of
Eq.(\ref{PBell}) (top);
the minimum-error probability of the POVM measurement of the field of Eq.(\ref{EMin}) (middle);
the fidelity of the postselected
Bell state $|\Psi^+\rangle_{AB}$ of Eq.(\ref{OptFid}) (bottom).}
\end{center}
\end{figure}

\begin{figure}[t]
\begin{center}
\includegraphics[width=8cm]{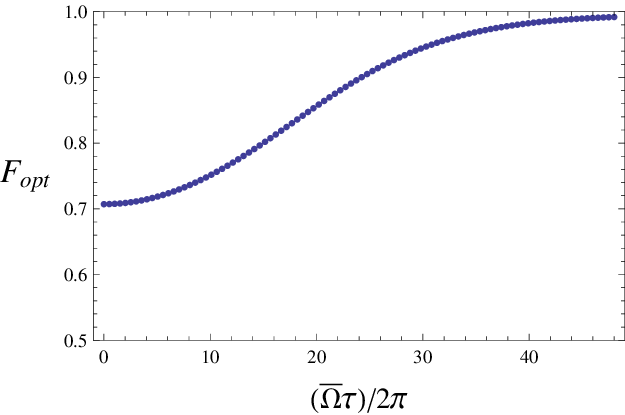}
\caption{\label{Fig3} Fidelity of the postselected
Bell state $|\Psi^+\rangle_{AB}$ of Eq.(\ref{OptFid}) in
the weak coupling limit $\Delta = 5 \overline{\Omega}_0$ for long interaction times: The effects of dephasing
lead to an asymptotic increase to unity}
\end{center}
\end{figure}

In Figs.~\ref{Fig1},\ref{Fig2}, and \ref{Fig3} numerical results are presented for characteristic
 quantitative measures which
exhibit to what extent postselection by a minimum-error POVM measurement on the optical radiation field is capable of
preparing a Bell state $|\Psi^+\rangle_{AB}$. These numerical results are based on the exact quantum state of
Eq.(\ref{Psi}) from which the minimum-error POVM measurement is determined according to Eq.(\ref{optimalPOVM}). This optimum
minimum-error POVM depends on the characteristic electrodynamical interaction parameters involved, namely the
interaction time $\tau$, the mean photon number $\overline{n}$, the detuning from resonance $\Delta$, and the strength
of the quantum electrodynamical coupling as measured by the resonant mean Rabi frequency $\overline{\Omega}_0$. In all
these figures the mean photon number of the initially prepared coherent field state $|\alpha\rangle$ is given by $\overline{n} =
|\alpha|^2 = 100$ so that typical quantum electrodynamical effects originating from photon number fluctuations as
measured by $\Delta n = \sqrt{\overline{n}} = 10$ are still apparent. In particular, this implies that deviations from 
the previously discussed analytical predictions of the linearization approximation are still observable.

Characteristic features of the strong coupling regime, i.e. $\Delta =0$, are depicted in Fig.\ref{Fig1} as a 
function of the dimensionless parameter $\overline{\Omega}\tau/(2\pi)$ which involves
the interaction time $\tau$ and the effective mean Rabi frequency $\overline{\Omega}$. 
Consistent with the
approximate analytical expression for the a priori probability $p$ as given by Eq.(\ref{apriori})
for shorter interaction times $\tau$ the success probability
$P_{Bell}$ of Eq.(\ref{PBell})
exhibits maxima at integer multiples of the interaction time $\tau = \pi/\overline{\Omega}$ (top). These
maxima correspond to multiples of Rabi cycles at which the material three-level systems 
are found again in their initially prepared
states $|0\rangle$ and $|1\rangle$ which constitute the qubits to be entangled. Correspondingly.
also minima appear at odd integer multiples of the interaction time $\tau = (\pi/2)/\overline{\Omega}$ at which the
three-level systems involved populate the excited state $|2\rangle$. However, for larger interaction
times these maxima and minima become less pronounced. This reflects the influence of the dephasing originating
from the dependence of the Rabi frequencies $\Omega(n)$ on the photon number $n$ in Eq.(\ref{Psi}) which is
also responsible for the collapse phenomena in the Jaynes-Cummings-Paul model \cite{Schleich}. 
In the limit of large interaction times, i.e.
$\overline{\Omega}\tau \geq \sqrt{\overline{n}}$ the success probability approaches the value 
$1/4$ consistent
with Eq.(\ref{apriori}).
The minimum-error probability of Eq.(\ref{EMin}) always exhibits maxima at completed
Rabi cycles, i.e. at integer multiples of the interaction time $\tau = \pi/\overline{\Omega}$ (middle).
Thus, 
at these interaction times 
it is difficult to distinguish the field states $\hat{\rho}_1$ and $\hat{\rho}_2$ even by a minimum-error
POVM measurement. This is due to the fact that in view of the periodic Rabi oscillations
at these interaction times the atom-field state is similar to the initially prepared quantum state
of Eq.(\ref{initialstate}) which is characterized by the property that the field states
$\hat{\rho}_1$ and $\hat{\rho}_2$ are identical and thus indistinguishable. This is also the reason 
for the vanishing
success probability at $\tau =0$. However, due to dephasing the difference between minima and maxima of 
the
minimum-error probability eventually vanishes in the limit of sufficiently large interaction times.
Similarly, also the fidelity of a Bell pair which is postselected by such a minimum-error POVM field measurement
exhibits periodic oscillations with the Rabi frequency $\overline{\Omega}$ (bottom). 
The maxima of these oscillations
appear at odd integer multiples of the interaction time $\tau = (\pi/2)/\overline{\Omega}$. At these interaction times
the three-level systems are likely to be found in the excited states $|2\rangle$ so that we expect
small success probabilities at these interaction times. However, due to dephasing and the related 
collapse phenomena 
at
sufficiently long interaction times
the differences
between minima and maxima of the fidelity eventually tend to zero and the achievable fidelity approaches 
its maximum possible value of unity. These sufficiently long interaction times of the order of
$\tau \geq \sqrt{\overline{n}}/\overline{\Omega}_0 = 1/\Omega_{vac}$
are therefore most favorable for
preparing a high-fidelity material Bell pair in the Bell state $|\Psi^+\rangle_{AB}$ provided
spontaneous emission of photons into other modes of the radiation field is negligible.  
Recent quantum electrodynamical experiments performed in the strong coupling regime \cite{Kimble}
demonstrate that such large vacuum Rabi frequencies are within current experimental possibilities.

This strong coupling behavior is in marked contrast to the dependence of these characteristic quantities
on the interaction time in the weak coupling limit in which the detuning is large, 
i.e. $|\Delta| \gg \overline{\Omega}_0$. This case is depicted in Figs.\ref{Fig2} and \ref{Fig3}. From the interaction times shown in Fig.\ref{Fig2} characteristic oscillations of these quantities
with the
mean Rabi frequency $\overline{\Omega} \to |\Delta/2|$ are apparent. They originate from the 
instantaneous turning on and off of the interactions between the optical radiation field and the
quantum systems $A$ and $B$. Although at their maxima the success probabilities $P_{Bell}$ are 
slightly larger than in the resonant case of Fig.\ref{Fig1} the fidelity of the postselected Bell pairs
is significantly smaller and oscillates slightly around the value of $F_{opt} = 1/\sqrt{2}$. In view
of the large detuning considered effects of dephasing are negligible for
the interaction times depicted in Fig.\ref{Fig2}. Effects of dephasing become important only in the limit of
extremely long interaction times of
the order of
$\tau \geq |\Delta/\overline{\Omega}_0|/\Omega_{vac} \gg 1/\Omega_{vac}$. The resulting collapse phenomena
cause an increase of the achievable fidelities so that asymptotically
they approach the maximum possible value of unity (compare with Fig.\ref{Fig3}).
However, typically at these extremely long
interaction times the 
influence of spontaneous emission of photons into other modes of the radiation field is no longer negligible
and our theoretical model is no longer adequate for describing such cases. Therefore, for the preparation
of high-fidelity Bell states by postselection the weak coupling regime exhibits clear limitations
even if the postselection is performed by minimum-error POVM measurements.

\section{Effects of photon loss\label{realistic}}
In this section we investigate additional 
effects of photon loss during the propagation 
of the optical radiation field through the optical fiber
by
modeling the propagation-induced photon loss by the
dynamics of damped harmonic oscillators with equal decay rates.
This may describe a physical situation, for example, in which
only a single transverse but many longitudinal modes of the radiation
field are excited in the optical fiber during the photonic
propagation process
and in which the
photon loss is due to leakage of this transversal mode
out of the optical fiber.
It is demonstrated that in the strong coupling
regime of resonant interactions besides an overall decay of success probabilities and achievable
fidelities of the postselected Bell pairs also characteristic
interference oscillations appear which originate from dephasing.

Let us consider again the quantum electrodynamical model of 
Sec.\ref{QED} with the only difference that
during the propagation of the optical radiation 
field through the fiber the dynamics 
is described by damped harmonic oscillators. 
Thus, during the time interval $T - 2\tau$ 
(with $\tau \ll 2/\Gamma \ll T$)
of the propagation through the optical fiber
the free dynamics of the optical radiation field 
previously described by the Hamiltonian 
$\sum_{i\in L} \hbar \omega_i \hat{a}_i^\dagger \hat{a}_i$
(compare with Appendix A)
is replaced by the Lindbladian master equation
\begin{equation}
\label{damp}
 \frac{d\hat{\rho}}{dt}=-i \sum_{i \in L} \omega_i [\hat{a}_i^\dagger 
\hat{a}_i,\hat{\rho}] + \sum_{i\in L} \left(
 [\hat{L}_i,\hat{\rho}\hat{L}^{\dagger}_i]
+ [\hat{L}_i\hat{\rho},\hat{L}^{\dagger}_i]\right)
=\mathcal{L}\hat{\rho}
\end{equation}
for the field state $\hat{\rho}(t)$
with the Lindblad operators 
$\hat{L}_i = \sqrt{\gamma/2}\hat{a}_i$.
Thereby, the damping rate
$\gamma$ characterizes the photon loss in
the optical fiber.

Expanding the density operator of a particular mode, say $i$, 
into photon number states, i.e. 
\begin{equation}
\hat{\rho}^{(i)} (t) =
\sum^{\infty}_{n,m=0}
 \rho^{(i)}_{n,m}(t)
{\rm exp}[-i\omega_i(n-m)t]\ket{n}_{ii}\bra{m}.
\end{equation}
and
taking matrix elements of the density operator equation 
\eqref{damp} we obtain the result
\begin{equation}
\dot{\rho}^{(i)}_{n,m}=-\frac{\gamma}{2} (m+n) 
 \rho^{(i)}_{n,m} + 
\gamma \sqrt{(n+1)(m+1)}  \rho^{(i)}_{n+1,m+1}. 
\end{equation}
A general solution of this equation 
can be obtained with the help of 
the Laplace transformation
\begin{equation}
\hat{\rho}^{(i)}(z) :=\int^{\infty}_0 \hat{\rho}^{(i)} (t) \ee^{-zt}\,dt 
\end{equation}
which transforms the master equation with the initial condition
$\hat{\rho}^{(i)}(t=0)$ into an algebraic equation for 
$\hat{\rho}^{(i)}(z)$.
Inverting its solution with the help of the inverse relation
\begin{equation}
\hat{\rho}^{(i)} (t)= \frac{1}{2 \pi i} 
\int_{\mathcal{C}} \ee^{zt} \hat{\rho}^{(i)} (z)\, dz  
\end{equation}
we obtain the corresponding solution $\hat{\rho}^{(i)}(t)$ for 
the field state of mode $i$ at time $t$. 
Thereby, the path of integration $\mathcal{C}$ has to 
be chosen in such a way that all poles of $\hat{\rho}^{(i)}(z)$
are included.

In the photon number-state representation
the time dependent solution of the master equation (\ref{damp}) 
can be determined by induction thus yielding the result
\begin{eqnarray}
\label{tsol}
\rho^{(i)}_{n,m}(t)&=&\ee^{-\gamma (m+n)t/2}
\sum^{\infty}_{j=0} \rho^{(i)}_{n+j,m+j}(t=0) \times 
\nonumber \\
&\times&\frac{\sqrt{(n+j)!}\sqrt{(m+j)!}}{\sqrt{n!}\sqrt{m!}}
\frac{(1-\ee^{-\gamma t})^j}{j!}
\end{eqnarray}
for mode $i$.
With the help of this solution it is straightforward to 
propagate field coherences of the form
 $|n\rangle_{ii} \langle m|$ of any mode $i$
from an initial time immediately after the excitation
by the optical cavity
field $A$ 
to the time 
after completion of the propagation through the optical fiber. 
Thus, an initial coherence between coherent states, such
as
$\ket{\beta}_{ii}\bra{\alpha}$, for example,
evolves to a coherence of the form
\begin{eqnarray}
\label{Lindbladsol}
&&\ee^{\mathcal{L}T} \ket{\beta}_{ii}\bra{\alpha}=\\ 
&&\ee^{-(1 - {\rm exp}(-\gamma T))
(\frac{|\alpha|^2+|\beta|^2}{2}-\beta \alpha^*)}
\ket{\beta\ee^{-\gamma T/2}}_{ii}\bra{\alpha \ee^{-\gamma T/2}}.\nonumber
\end{eqnarray}

\begin{figure}[t]
\begin{center} 
\includegraphics[width=8.cm]{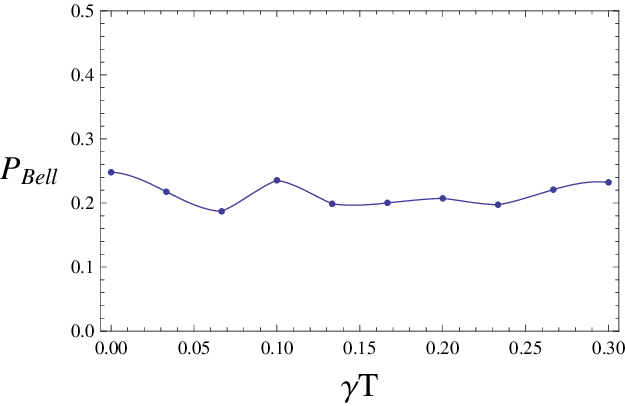}
\includegraphics[width=8cm]{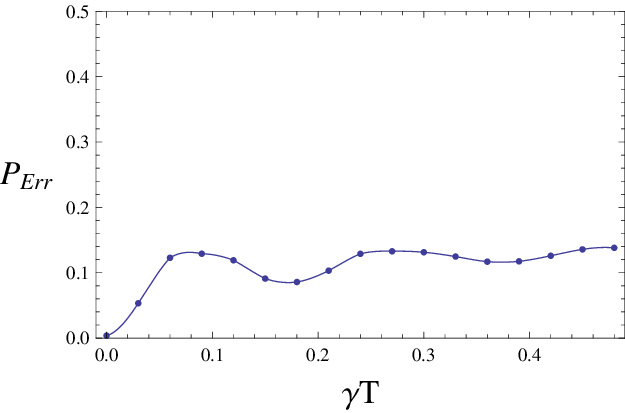}
\includegraphics[width=8cm]{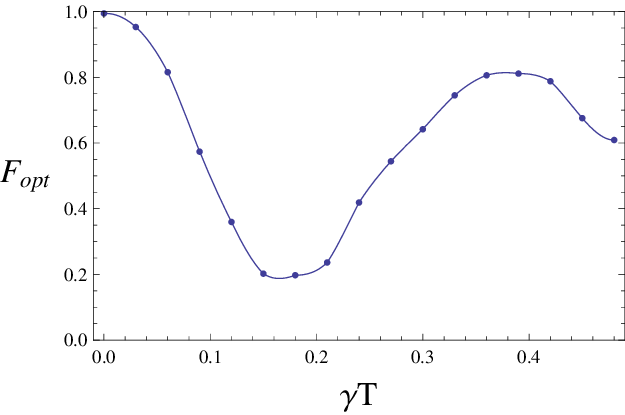}
\caption{\label{Fig4}Dependence of characteristic quantities on 
the propagation time $T$ ( in units of $1/\gamma$) through a lossy optical fiber in
the strong coupling limit $\Delta = 0$ for an interaction time $\tau = (23/4) 2\pi/\overline{\Omega}_0$:
The success probability for postselecting a Bell state 
$|\Psi^+\rangle_{AB}$ of 
Eq.(\ref{PBell}) (top);
the minimum-error probability of the POVM measurement of the field of Eq.(\ref{EMin}) (middle); 
the fidelity of the postselected
Bell state $|\Psi^+\rangle_{AB}$ of Eq.(\ref{OptFid}) (bottom).
} 
\end{center}
\end{figure}

In Fig.\ref{Fig4} numerical results are presented which demonstrate characteristic properties
of the postselection of a Bell state $|\Psi^+\rangle_{AB}$ by a minimum-error POVM measurement of
the optical radiation field in the presence of photon loss during propagation through the optical fiber.
Thereby,
the quantum state resulting from all interactions between the optical field and the quantum
systems $A$ and $B$ has been evaluated numerically with the help of relation (\ref{tsol}). 
Apart from the photon loss during the propagation through the optical fiber 
and the choice of a fixed interaction time the parameters are
the same as in Fig.\ref{Fig1}. The interaction time $\tau = (23/4)2\pi/\overline{\Omega}_0$
has been chosen in such a way that
in the absence of photon loss
the fidelity of creating a Bell state has a maximum and that
its value is close to unity (compare with Fig.\ref{Fig1}).
Fig.\ref{Fig4} depicts the dependence of the characteristic quantities $P_{Bell}$, $E_{min}$, and $F_{opt}$
of this minimum-error postselection procedure on the time $T$
of propagation through the fiber. It is apparent that photon loss tends to decrease the success probability
$P_{Bell}$ and the fidelity $F_{opt}$ of the postselected Bell pair. 
At the same time it also
increases the minimum error $E_{min}$. For an  optical fiber with an intensity loss of 
$ D~ {\rm dB}/{\rm m}$ propagation durations $T$ can be translated into lengths $L$ 
of the optical fiber by the relation $L = (\gamma T) 20 /(D {\rm ln}10)$.
Thus, at a wave length of $1550 {\rm nm}$
with a photon loss of $0.2 {\rm dB}/{\rm km}$, for example, the maximum value of $\gamma T = 0.3$
depicted in Fig.\ref{Fig4} corresponds to a fiber length of $L =  13029 {\rm m}$.
It is interesting to note that all the characteristic quantities depicted in Fig.\ref{Fig4}
exhibit also
an oscillatory behavior. It can be traced back to the fact that according to Eq.(\ref{Lindbladsol})
for small photon loss, i.e. $\gamma T \ll 1$, all relevant field coherences
between coherent states also involve a characteristic frequency 
$\tilde{\omega} = \gamma {\rm Im}(\alpha^* \beta)$. According to Eq.(\ref{coherent})
due to dephasing typical relevant coherent states
are of the form $|\alpha e^{ik\theta}\rangle$ with $k=\pm 1,\pm 2,...$. Therefore, an estimate of these
characteristic frequencies can be obtained by assuming that $\alpha = \sqrt{\overline{n}} e^{i\theta}$
and $\beta = \sqrt{\overline{n}}$, for example, which yields an oscillation frequency of the order of
$\tilde{\omega} = 
\gamma \overline{n}(\overline{\Omega}_0\tau/(2\overline{n})) = \overline{\Omega}_0 \gamma\tau/2$
for resonant coupling , i.e. $\Delta =0$.

\section{Summary and Conclusion}
\label{DaC}

In the context of the hybrid quantum repeater model 
we have studied optimal
possibilities of preparing high-fidelity Bell pairs of two material qubits by postselection
with the help of two single-mode cavities and an optical fiber connecting them.
Whereas the original proposal of van Loock et al. 
\cite{vanLoock1}
concentrated on the dynamical regime of weak coupling in which the interaction between the optical radiation
field and the two material quantum systems involved can be described perturbatively, our discussion 
concentrated on
the strong coupling regime of almost resonant interaction. 
We also discussed the problem of 
photonic quantum state transfer through an optical fiber 
and determined
by what choices of the coupling parameters between cavities and
optical fiber perfect photonic quantum state transfer is possible.
We determined the optimum POVM measurements which have to be performed on 
the single-mode optical radiation field of cavity $B$
in order to postselect a Bell pair with minimum error. On the basis of this analysis we demonstrated that
some of the limitations of the non-resonant weak coupling limit can be circumvented successfully in the strong
coupling regime. In particular, provided the propagation of the optical radiation field through the fiber is ideal
it is possible to create Bell pairs of fidelities arbitrarily close to unity provided the interaction time
between the material quantum systems and the optical field is chosen properly. This is due to the fact that
at these particular interaction times the quantum states of the optical radiation field which are entangled
with the desired two-qubit Bell state and with the other material quantum states involved are almost orthogonal
so that they can be distinguished almost perfectly by a von Neumann measurement.
According to Eq.(\ref{coherent}) this von Neumann measurement
involves an approximate projection onto a cat-state-like superposition of two coherent states.
This is in marked contrast
to the weak coupling regime of non-resonant interaction where the relevant field states are always non orthogonal
so that they cannot be distinguished perfectly by any quantum measurement and the resulting postselection
is never perfect. Furthermore, due to photon-induced dephasing effects which are characteristic for
the collapse phenomena of the Jaynes-Cummings-Paul model
at these interaction times the success probabilities 
of the strong coupling regime tend to the limiting value of $1/4$ and are
not significantly smaller than the corresponding values (of the order of $0.4$)
achievable in the weak coupling regime. We have also explored effects originating from photon
loss taking place during the propagation of the optical radiation field through the optical fiber. In addition
to an overall decrease of achievable fidelities and success probabilities, in the strong coupling regime also
an oscillatory behavior is apparent.
Thus, the strong coupling regime of the hybrid quantum repeater
model offers interesting
perspectives for entangling material quantum systems over not too long distances and thus for providing
high-fidelity Bell pairs at high repetition rates for future realizations of quantum repeaters.

\section{Acknowledgment}

This work is supported by the BMBF project QK{\_}QuOReP. 
We would like to thank Peter van Loock, Gerd Leuchs, and Thomas Walther for helpful discussions.

\appendix
\section{Perfect quantum state transfer between two single-mode cavities coupled by an optical fiber}
\label{Appendix}

In this Appendix it is demonstrated that within the validity of the
rotating-wave and pole approximation perfect quantum state 
transfer
between two single modes of two spatially well separated cavities connected by
a long optical fiber is possible provided
the coupling between the cavities and the optical fiber
is engineered appropriately. Explicit expressions for these optimal couplings
are determined.

In the following we consider a quantum state transfer scenario in which
a single mode of cavity $A$ 
with frequency $\omega$
is coupled 
almost resonantly
to a dense set of modes of an optical fiber. Initially, a photonic quantum state is prepared in this single mode
of cavity $A$ and all other modes of the fiber and of the second distant cavity $B$
are prepared in their vacuum states.
After having left cavity $A$ the photons propagate through the fiber so that
cavity $A$ is left in its vacuum state (apart from exponentially small terms).
After this decay process a photonic wave packet propagates through the optical fiber
whose spatial extension is of the order of $c/\Gamma_A$ with $c$ denoting the speed of photonic
propagation in the fiber and with $\Gamma_A$ denoting the resonant decay rate of cavity $A$ into the fiber.
In particular, we assume that the length $L$ of the optical fiber is large, i.e. $L\gg c/\Gamma_A$.
Thus, during the propagation of the photon wave packet through the optical fiber the coupling of the fiber modes
to the second field mode of frequency $\omega$ which is localized in the second distant cavity $B$
is turned
on adiabatically. In the following it is shown that
the photonic wave packet can leak from the optical fiber into cavity $B$ almost perfectly provided
the couplings between both cavities and the optical fiber are engineered appropriately.

In the rotating wave approximation the almost resonant coupling of the single mode 
of frequency $\omega$ in cavity $A$ to the
optical fiber modes is described by the Hamiltonian
\begin{eqnarray}
\hat{H}&=& \hbar \omega \hat{a}^\dagger_A\hat{a}_A +
\sum_{i\in L} \hbar \omega_i \hat{a}^\dagger_i\hat{a}_i + 
\sum_{i \in L} \left( \kappa_i 
\hat{a}^\dagger_i \hat{a}_A+ \kappa^*_i 
\hat{a}^\dagger_A \hat{a}_i\right) = \nonumber \\
&& \sum_{k,j\in I}\hat{a}^\dagger_k H_{kj} \hat{a}_j
\label{Hamiltonian10}
\end{eqnarray}
with $L$ denoting the set of modes of the optical fiber
and with $I = L \cup \{A\}$. 
As we want to transfer a photonic quantum state through a long
optical fiber we assume in our subsequent discussion that only one transversal mode
but a large number of longitudinal modes of the optical fiber couple to the optical cavity $A$
almost resonantly in the frequency band $(-\Delta + \omega, \omega + \Delta)$. The rotating wave
approximation is valid if $\Delta \ll \omega$.
In a simple approximation one may assume that
in this frequency band the frequencies of the almost resonantly 
coupled modes of the optical fiber can be described by the relation
$\omega_n = 2\pi c n/l$
 with integer values of $n$ and with $l$ denoting the length of the optical fiber.
($c$ is the speed of light inside the optical fiber.)
Accordingly, in terms of the bare modes the matrix representation
of the Hamiltonian of Eq.(\ref{Hamiltonian10}) is given by
\begin{eqnarray}
 \hat{H} &\longrightarrow&\left( \begin{array}{ccccc}
\hbar \omega & \kappa^*_1 & \kappa^*_2 & . & .\\
\kappa_1 & \hbar \omega_1 & 0 & . & .\\
\kappa_2 & 0 & \hbar \omega_2 & . & .\\
. &  . &  . & .& . \\
. &  . &  . & .& .\end{array} \right). 
\end{eqnarray}
This Hamiltonian can be diagonalized by
a unitary transformation $\hat{U}$
which describes the transformation
to normal coordinates or to dressed modes, i.e.
\begin{eqnarray}
 \hat{H} &=&\hat{U} \Lambda \hat{U}^{\dagger}, \, \, \,
 \Lambda~\longrightarrow~\left( \begin{array}{ccccc}
\hbar \lambda_0 & 0 & 0 & . & .\\
0 & \hbar \lambda_1 & 0 & . & .\\
0 & 0 & \hbar \lambda_2 & . & .\\
. &  . &  . & .& . \\
. &  . &  . & .& .\end{array} \right)
\end{eqnarray}
with the dressed eigenfrequencies 
$\{\lambda_j\}~(j\in {\mathbb N}_0)$ and the destruction
and creation operators of the dressed eigenmodes
$\hat{a}'_j = \sum_{i\in I}
\hat{U}^\dagger_{ji} \hat{a}_i$ and
$\hat{a}'^\dagger_j =
\sum_{i\in I} U_{ji}\hat{a}^\dagger_i$, 
respectively. Thus, in these dressed modes the Hamiltonian
is diagonal, i.e.
\begin{equation}
\hat{H}=\sum_{j \in {\mathbb N}_0} 
\hbar \lambda_j \hat{a}'^\dagger_j\hat{a}'_j.
\end{equation}
For $i\in L$ and $j \in {\mathbb N}_0$
the matrix elements of the diagonalizing unitary transformation $\hat{U}$
are given by
\begin{eqnarray}
U_{ij}&=&-\frac{\kappa_i}{\hbar \omega_i - \hbar \lambda_j} U_{Aj} := U_i(\lambda_j),\nonumber\\
U_{Aj} &=& \frac{1}{\sqrt{1 + \sum_{n \in L}\frac{|\kappa_n|^2}{(\hbar \omega_n-\hbar \lambda_j)^2}}}:= U_A(\lambda_j)
\end{eqnarray}
with the dressed frequencies $\{\lambda_j \}$
being determined by the zeros of the quantization 
function $f(\lambda)$, i.e.
\begin{eqnarray}
 f(\lambda_j) &:=&
\hbar \omega - \hbar \lambda_j
-\sum_{n\in L}
\frac{|\kappa_n|^2}{\hbar \omega_n-\hbar \lambda_j} = 0.
\label{quanizationfunction}
\end{eqnarray}

With the help of this transformation to normal coordinates 
the time evolution of any coherent initial  state 
$\ket{\Psi(t=0)}=\ket{\{\alpha_i\}}=\ket{\alpha_0}_{A}\otimes
\ket{\alpha_1}_1 \otimes\ket{\alpha_2}_2.....$ 
of the bare modes with $\hat{a}_i|\beta\rangle_i = \beta |\beta\rangle_i$ and
$i\in I$
can be determined easily because for $t\geq 0$
\begin{eqnarray}
 e^{-it\hat{H}/\hbar}\ket{\{\alpha_i\}}=
e^{-\sum_{i\in I}|\alpha_i|^2/2} 
e^{\sum_{i\in I} \hat{a}^\dagger_i \alpha_i(t)}\ket{0} \equiv
\ket{\{\alpha_i(t)\}}\nonumber\\
&&
\label{timeevolution1}
 \end{eqnarray}
with the vacuum state of all modes $|0\rangle$ and with ($i\in I$)
\begin{eqnarray}
\label{multisol}
\alpha_i(t)&=&\sum_{k\in {\mathbb N}_0,j\in I} 
U_{ik} 
e^{-i \lambda_{k} t} U^*_{jk} \alpha_j.
\end{eqnarray}
Thereby, we have taken into account the identity
\begin{equation}
 \sum_{j \in I} |\alpha_j(t)|^2 =\sum_{j\in I} |\alpha_j|^2 
\end{equation}
which is a consequence of the unitary time evolution.
The time evolution of $\alpha_i(t)~(i\in I)$ 
as given by
Eq.(\ref{timeevolution1})
can also be  determined
by complex integration, i.e.
\begin{eqnarray}
\alpha_i(t)&=&
\frac{-1}{2 \pi i}
\int_{-\infty +i0}^{\infty+i0} d\lambda
U_i(\lambda) \frac{df/d\lambda(\lambda)}{f(\lambda)}  \sum_{m \in I} \alpha_m U^*_m(\lambda) e^{-i\lambda t}.\nonumber\\
\label{Laplace10}
\end{eqnarray}
Evaluating Eq.(\ref{Laplace10}) with the help of the calculus of residues
one arrives at Eq.(\ref{timeevolution1}).

In the continuum limit of a very long optical fiber, in which the bare mode frequencies $\{\omega_i, i\in I\}$
are so densely spaced that they
can be described by a continuum, the quantization function of 
Eq.(\ref{quanizationfunction}) can be approximated by
\begin{eqnarray}
f(\lambda + i0)&=&\hbar (\omega - \delta \omega) - \hbar \lambda-i\frac{\hbar \Gamma_A}{2}
\label{poleapproximation}
\end{eqnarray}
with
\begin{eqnarray}
\Gamma_A&=&\frac{2\pi}{\hbar}|\kappa_n|^2
\frac{dn}{d\hbar \omega_n}\biggl|_{\omega_n=\omega}.
\end{eqnarray}
This approximation 
is an immediate consequence of the relation
\begin{equation}
\sum_{n \in L}\frac{|\kappa_n|^2}{\hbar \omega_n-\hbar \lambda -i0}=
\hbar \delta \omega + i\frac{\hbar \Gamma_A}{2}
\end{equation}
and of the
Weisskopf-Wigner or pole 
approximation \cite{Wigner-Weisskop}. This latter approximation assumes that
the interaction-induced frequency shift $\delta \omega$ as well as the decay rate $\Gamma_A$ are slowly varying functions
of $\lambda$ so that they can be replaced by their values at $\lambda = \omega$ and can be considered
as being $\lambda$-independent. For its validity it is necessary that both the frequency shift $\delta \omega$ and the
decay rate $\Gamma_A$ are small in comparison with the cavity frequency $\omega$, i.e.
$|\delta \omega|, \Gamma_A \ll \Delta \ll \omega$.
In the following we absorb
the frequency shift $\delta \omega$ in a renormalized cavity frequency, i.e. $\omega -\delta \omega ~\to~\omega$.
The $\lambda$-independent decay rate
$\Gamma_A$
describes the loss of photons from cavity $A$ due to the coupling of the cavity mode with frequency $\omega$ to the
optical fiber.

From Eqs.(\ref{Laplace10}) and (\ref{poleapproximation}) the solution of the Schr\"odinger equation with Hamiltonian
(\ref{Hamiltonian10}) and with initial condition
\begin{equation}
\ket{\psi(t=0)}=\ket{\alpha}_{A}\prod_{i\in L}\ket{0}_i
\label{initialcondition}
\end{equation}
is given by
$|\psi(t)\rangle = |\alpha(t)\rangle_A \prod_{i\in L}\ket{\alpha_i(t)}_i$ with
\begin{eqnarray}
\label{depletionsol}
 \alpha(t)&=&\alpha \ee^{-i \omega t-\Gamma_A t/2}, \\
\alpha_i(t)&=&\frac{\alpha\kappa_i}{\hbar \omega_i-\hbar \omega + i \hbar \Gamma_A/2}
\left(\ee^{-i\omega_i t}- \ee^{-i \omega t-\Gamma_A t/2} \right). \nonumber
\end{eqnarray}
Thus,
for a sufficiently long interaction time, i.e. $T_1 \gg 1/\Gamma_A $, apart from exponentially 
small terms of the order of
${\rm exp}(-\Gamma_A T_1/2)$ 
the depletion of the cavity mode is perfect
and $|\psi(t)\rangle$ describes a pure quantum state in which the cavity mode $A$ is in its vacuum state and
a photonic wave packet propagates through the optical fiber. 

Let us now consider a time $T_1$ with $T_1\gg 1/\Gamma_A$ at which 
the occupied modes of the radiation field are described
by the quantum state $\prod_{i\in L}\ket{\alpha_i(T_1)}_i$ and at which cavity $A$ 
is approximately in its vacuum state. This quantum state describes a photonic wave packet in the
optical fiber with a spatial extension of the order of $c/\Gamma_A$. 
If the optical fiber is sufficiently large, i.e. $l \gg c/\Gamma_A$, at this time $T_1$
this photonic wave packet
is localized well inside the optical fiber and propagates towards the second cavity $B$ which is
assumed to be prepared in its vacuum state at time $T_1$. In the course of
this propagation process
the coupling between the optical fiber and cavity $B$ which is negligible at time $T_1$
is turned on adiabatically and eventually leads to the leakage of this photon wave packet into cavity $B$.
Thus, in analogy with Eq.(\ref{Hamiltonian10}) for times $t\geq T_1$
the interaction Hamiltonian between the optical fiber and this second cavity $B$ is given by
\begin{eqnarray}
\hat{H}'&=& \hbar \omega \hat{a}^\dagger_B\hat{a}_B +
\sum_{i\in L} \hbar \omega_i \hat{a}^\dagger_i\hat{a}_i + 
\sum_{i \in L} \left( \kappa'_i 
\hat{a}^\dagger_i \hat{a}_B+ \kappa^{'*}_i 
\hat{a}^\dagger_B \hat{a}_i\right).\nonumber\\
\label{Hamiltonian11}
\end{eqnarray}
with complex-valued coupling coefficients $\kappa'_i$. In general they may differ from the 
coupling coefficients $\kappa_i$ for cavity $A$.
Again, in the continuum limit and in the pole approximation one can characterize the coupling between
the optical fiber and cavity $B$ by a decay rate 
\begin{eqnarray}
\Gamma_B &=& 
\frac{2\pi}{\hbar}|\kappa'_n|^2
\frac{dn}{d\hbar \omega_n}\biggl|_{\omega_n=\omega}.
\end{eqnarray}
Evaluating the quantum state of the optical fiber and of cavity mode $B$
$|\psi(T_1+t)\rangle$ for $t\gg 1/\Gamma_B$ with the help of Eq.(\ref{Laplace10}) we obtain the result
\begin{eqnarray}
|\psi(T_1+t)\rangle &=&
|\alpha(T_1+t)\rangle_B\prod_{i\in L}|\alpha_i(T_1+t)\rangle_i
\end{eqnarray}
with
\begin{eqnarray}
\label{excitation}
&&\alpha(T_1+t) =
\sum_{i\in L} \frac{\kappa'^*_i}{\hbar \omega_i-\hbar \omega+ i\hbar \Gamma_B/2}
\frac{\alpha \kappa_i\ee^{-i\omega_i (T_1+t)}}{\hbar \omega_i-\hbar \omega+i \hbar \Gamma_A/2},\nonumber\\ 
&&\alpha_i(T_1+t) =\frac{1}{2 \pi i}\int_{-\infty +i0}^{\infty+i0} d\lambda~\frac{\kappa'_i}{\hbar \omega_i-\hbar \lambda}
\frac{e^{-i\lambda t}}{\omega-\lambda - i\Gamma_B/2} \times \nonumber \\
&&\times \sum_{j \in L}
\frac{\alpha \kappa_j\ee^{-i\omega_j T_1}}{\hbar \omega_j-\hbar \omega+i \hbar \Gamma_A/2} 
\frac {\kappa'^*_j}{\hbar \omega_j-\hbar \lambda},
\end{eqnarray}
where in the first equation we already neglected terms which are exponentially small in the parameter ${\rm exp}(-\Gamma_B t/2)$.
A comparison of Eq.(\ref{excitation}) with the initial condition of Eq.(\ref{initialcondition}) 
shows that
for interaction times $T_1+t$ with $t \gg 1/\Gamma_B$ the necessary and sufficient condition for perfect quantum
state transfer between cavities $A$ and $B$ is $\alpha = \alpha(T_1+t)$. A sufficient condition for satisfying this
condition
is to choose the coupling constants between the optical fiber and the second cavity $B$ in such a way that
\begin{eqnarray}
\kappa'_i &=& \kappa_i^* = \mid \kappa_i\mid e^{-i\varphi_i},\nonumber\\
e^{2i\varphi_i} &=& \frac{\hbar \omega_i - \hbar \omega + i\hbar \Gamma_A/2}
{\hbar \omega_i - \hbar \omega - i\hbar \Gamma_A/2}
\label{phases}
\end{eqnarray}
and to choose the interaction time $t = T_2$ with $T_1,T_2 \gg 1/\Gamma_A,1$
so that $\omega_n (T_1+T_2)$ 
is an integer multiple of $2\pi$. If the relevant
modes of the optical fiber fulfill the condition $\omega_n = 2\pi c n/l$ with integer values of $n$, 
for example,
this latter condition
can be fulfilled by the choice $c(T_1+T_2) = l$. 
This condition describes the fact that the total interaction time $T=T_1+T_2$ 
has to allow for a propagation of photons through the
optical fiber of length $l$. In addition, this interaction time has to be large enough 
so that the leaking out of cavity $A$ and the
leaking into cavity $B$ can be completed, i.e. $T_1,T_2 \gg 1/\Gamma_A$.
The complex conjugation involved in Eq.(\ref{phases}) reflects the fact that 
for perfect quantum state transfer a time
reversal process is necessary.
The phase modulation of the coupling constants $\kappa_i$
required by Eq.(\ref{phases}) is characteristic for the scattering phase shift 
of a Breit-Wigner resonance with a Lorentzian frequency distribution.


\begin{thebibliography}{99}
%
%
%
\bibitem{Leuchs}
{\it Quantum Information Processing}, edited by Th. Beth and G. Leuchs
(Wiley-VCH, Weinheim, 2005).
%
\bibitem{repeaterreview} N. Sangouard, C. Simon, H. de Riedmatten, and N. Gisin, Rev.
Mod. Phys. {\bf 83}, 33 (2011).
%
\bibitem{Briegel98} H.-J. Briegel, W. D\"ur, J. I. Cirac, and P. Zoller,
Phys. Rev. Lett. {\bf 81}, 5932 (1998).
%
\bibitem{Duer99} W. D\"{u}r, H.-J. Briegel, J. I. Cirac, and P. Zoller,
Phys. Rev. A {\bf 59}, 169 (1999).
%
\bibitem{vanLoock1} P. van Loock, T. D. Ladd, K. Sanaka, F. Yamaguchi, K. Nemoto, W. J. Munro, and Y. Yamamoto, 
Phys. Rev. Lett. {\bf 96}, 240501 (2006).
\bibitem{vanLoock2} T. D. Ladd, P. van Loock, K. Nemoto, W. J. Munro, and Y.
Yamamoto, New J. Phys. {\bf 8}, 184 (2006).
%
\bibitem{vanLoock3} P. van Loock, N. L\"utkenhaus, W. J. Munro, and K. Nemoto, Phys. Rev. A {\bf 78}, 062319 (2008).
%
\bibitem{Cirac} J. I. Cirac, P. Zoller, H. J. Kimble, and H. Mabuchi, Phys. Rev. Lett. {\bf 78}, 3221 (1997). 
%
\bibitem{Enk} S. J. van Enk, J. I. Cirac, and P. Zoller, Phys. Rev. Lett. {\bf 79}, 5178 (1997). 
%
\bibitem{Pellizzari} T. Pellizzari, Phys. Rev. Lett. {\bf 79}, 5242 (1997).
%
\bibitem{Steinmetz} T. Steinmetz, Y. Colombe, D. Hunger, T.W. H\"ansch, A. Balocchi, Appl. Phys. Lett. {\bf 89}, 111110 (2006).
%
\bibitem{Colombe}Y. Colombe, T. Steinmetz, G. Dubois, F. Linke, D. Hunger, J. Reichl, Nature (London) {\bf 450}, 272 (2007).
%
\bibitem{Schleich} W. P. Schleich, {\it Quantum Optics in Phase Space}
(Wiley-VCH, Weinheim, 2001).
%
\bibitem{Helstrom} C. W. Helstrom, {\it Quantum Detection and Estimation Theory} (Academic,
New York, 1976).
%
\bibitem{Holevo} A. S. Holevo, Trans. Moscow Math. Soc. {\bf 26}, 133 (1972).
%
\bibitem{Hayashy} M. Hayashi, {\it Quantum Information} (Springer, Berlin, 2006).
%
\bibitem{Chefles} A. Chefles, Contemp. Phys. {\bf 41}, 401 (2000).
%
\bibitem{Chefles2} A. Chefles, Phys. Lett. A {\bf 239}, 399 (1998).
%
\bibitem{Bruss} M. Kleinmann, H. Kampermann, and D. Bruss, Phys. Rev. A {\bf 81}, 020304 (2010).
%
\bibitem{NielsenChuang} M.A. Nielsen and A. L. Chuang, {\it Quantum Computation and Quantum Information} (Cambridge University Press,
 Cambridge, 2000).
%
\bibitem{Kimble} J. McKeever, A. Boca, A. D. Boozer, J. R. Buck, and H. J. Kimble, Nature (London) {\bf 425}, 268 (2003).
%
\bibitem{Wigner-Weisskop} V. Weisskopf and E. Wigner, Z. Phys. {\bf 63}, 54 (1930); Z. Phys. {\bf 65}, 18 (1930). 
%
%
%
%

%

%

%
%
%
%

%


\end{thebibliography}
\end{document}